# A Highly Sensitive Diamond NV Magnetometer Using Ramsey Interferometry with a Short Sensor-to-Sample Distance

Yuta Araki,[1] Takeharu Sekiguchi,[1] Yuji Hatano,[1] Naota Sekiguchi,[1] Chikara Shinei,[2,3] Masashi Miyakawa,[4] Takashi Taniguchi,[4] Tokuyuki Teraji,[2] Hiroshi Abe,[5] Shinobu Onoda,[5] Takeshi Ohshima,[5,6] Takayuki Shibata,[7] Mutsuko Hatano,[1] and Takayuki Iwasaki[1]

[1])Department of Electrical and Electronic Engineering, School of Engineering, Institute of Science Tokyo, Meguro, Tokyo 152-8552, Japan

[2])Research Center for Electronic and Optical Materials, National Institute for Materials Science, Tsukuba, Ibaraki 305-0044, Japan

[3])Department of Applied Physics, University of Tsukuba, Tsukuba, Ibaraki 305-8571, Japan

[4])Research Center for Materials Nanoarchitectonics, National Institute for Materials Science, Tsukuba, Ibaraki 305-0044, Japan

[5])Takasaki Institute for Advanced Quantum Science, National Institutes for Quantum Science and Technology : QST, Takasaki, Gunma 370-1292, Japan

[6])Department of Materials Science, Tohoku University, Sendai, Miyagi 980-8579, Japan

[7])Advanced Research and Innovation Center, DENSO CORPORATION, 500-1, Minamiyama Komenoki, Nisshin, Aichi 470-0111, Japan

(*Electronic mail: iwasaki.t.aj@m.titech.ac.jp )

In this study, we developed a diamond quantum magnetometer based on Ramsey interferometry with a short sensor-to-sample distance. Conventional biomagnetic sensors with ensemble nitrogen-vacancy centers using continuous-wave optically detected magnetic resonance and Ramsey methods typically rely on watt-level lasers to achieve high sensitivity, resulting in thermal issues. In contrast, by employing the light-trapping diamond waveguide technique in a high-pressure and high-temperature diamond sample treated with electron beam irradiation, we obtained a high photon conversion efficiency of 9.5%, enabling us to simultaneously achieve a high sensitivity of 2.93(7) pT/$\sqrt{\text{Hz}}$ in the 100-400 Hz frequency range and a minimal temperature increase of only approximately 13 K at a low laser power of 210 mW. Using a dry phantom designed to mimic magnetoencephalography signals, we measured a weak magnetic field of 77.7(2) pT without signal averaging at a sensor-to-sample distance of 2.5 mm. This short-distance measurement prevents severe spatial signal attenuation, yielding a high signal-to-noise ratio. The development here is crucial for practical biomagnetic applications based on Ramsey interferometry.

Nitrogen-vacancy (NV) centers in diamond have the potential to measure biomagnetic fields with high sensitivity and spatial resolution at room temperature.[1] Previous studies have measured biomagnetic fields such as action potentials of marine worms and squid,[2] cardiac signals in rats,[3,4] and signal from mouse muscle[5] with ensemble NV centers using continuous-wave optically detected magnetic resonance (CW-ODMR). However, further improvement in magnetic-field sensitivity is required to detect much weaker signals such as those of interest in magnetoencephalography (MEG). Theoretically, the Ramsey method, which uses spin coherence between two energy states, provides a higher sensitivity than CW-ODMR by avoiding microwave (MW) and optical power broadening.[6] A recent study[7] has reported a high sensitivity of 0.46 pT/$\sqrt{\text{Hz}}$ (>2 kHz) based on the Ramsey method. Their sensor configuration optimized solely for sensitivity relied on watt-level optical excitation (4 W) and bulky MW and radio-frequency components. These restrictions raised the sensor temperature by 60 K and prevented the sensor from approaching a sample within 6.6 mm. Since biomagnetic fields decay steeply with distance and excessive heat causes invasive effects on biological tissues, overcoming these fundamental thermal and geometrical issues is essential for millimeter-scale biomagnetic field sensing.

In this study, we developed a Ramsey-based diamond quantum sensor operating at a low laser power. By employing the light-trapping diamond waveguide technique in an electron-beam-irradiated high-pressure and high-temperature (HPHT) diamond, we achieved a 9.5% photon conversion efficiency. We utilized a printed-circuit-board (PCB) MW antenna to achieve a short sensor-to-sample distance, and operated at a low laser power of 210 mW. This approach suppressed the rise of the diamond temperature to only ~13 K, safely maintaining the temperature below the 42 °C threshold for biological safety,[8] allowing for a short sensor-to-sample distance of just 2.0 mm. Under this practical configuration, the system exhibited a magnetic sensitivity of 2.93(7) pT/$\sqrt{\text{Hz}}$ in a frequency range of 100–400 Hz. We also investigated the sensing capability at a short sensor-to-sample distance of 2.5 mm using a dry phantom.

Figure 1 (a) shows a schematic of the entire sensor setup housed within a three-layer magnetic shield. A sensor head was centered between a pair of circular arrays[7,9] of sixteen Sm-Co magnets to lift the spin level degeneracy of $|m_s = \pm 1\rangle$. A bias magnetic field of 0.52 mT was applied along the [111] axis of the NV center. The axial distance between the magnet arrays and the diamond was optimized to minimize the inhomogeneity of the magnetic field in the diamond. The magnet arrays provided apertures with a diameter of 130 mm and an open space with a length of 137 mm. This configuration allowed a small animal such as a rat to be placed close to the sensor.

As shown in Fig. 1 (b), the sensor head comprised a diamond with an ensemble of NV centers, a dielectric mirror to reflect the fluorescence, a photon collection system, and an MW antenna. The shortest sensor-to-sample distance was 2.0 mm, limited by the thickness of the antenna and the dielectric mirror. The diamond was synthesized using a HPHT method with a $^{12}$C-enriched carbon source (99.998% isotopic purity). The $^{12}$C purity in the resulting HPHT crystal was 99.995%, corresponding to [$^{13}$C] = 50 ppm.[10] The nitrogen concentration was controlled by adding nitrogen getters of appropriate concentration.[11] The concentrations of P1 and NV$^-$ centers after electron-beam irradiation (EBI) were estimated by electron spin resonance as 1.2(1) ppm and 0.20(4) ppm, respectively. The thickness of the diamond was 0.35 mm and its area was approximately 0.9 mm × 0.9 mm, featuring a 60° angled facet at one edge. The surface of the diamond was oriented in the [111] direction. As shown in Fig. 1(c), a laser beam with a wavelength of 532 nm and a power of 210 mW was focused onto the diamond sample using a convex lens with a focal length of 200 mm. The focused beam had a diameter of approximately 30 $\mu$m and a Rayleigh length of about 1.3 mm. The fluorescence was collected by a borosilicate-glass TIR lens to enhance the collection efficiency[6,12] and filtered by a 650 nm long-pass filter and a 900 nm short-pass filter. To enhance the fluorescence intensity from the NV centers under low laser power, we employed the light-trapping diamond waveguide technique.[13] Total internal reflection in the diamond was clearly observed in Fig. 1(b), which extended the path length to 8.8 mm, greater than the single path length of 0.9 mm by a factor of roughly 10. This configuration yielded approximately 20 mW of collected fluorescence from an incident laser power of ~210 mW, which corresponds to a photon conversion efficiency of 9.5%. This high efficiency allows for operation at an incident laser power an order of magnitude lower than that typically used in biomagnetic field measurements[2–5] and Ramsey measurement.[7] Pulsed-ODMR measurements confirmed that this low-power operation restricted the temperature rise to only ~13 K, resulting in the diamond temperature of approximately 38 °C. This remains safely below the 42 °C biological heat-pain threshold,[8] resolving the thermal issues for non-invasive sensing while preserving sufficient signal intensity.

We employed a balanced detection technique[14,15] to reduce laser intensity noise and applied lock-in detection combined with pulse measurements to reach the photon shot noise limit. The laser beam was split into two paths using a 70:30 nonpolarizing beam splitter (NPBS). As shown in Fig. 1(c), the reference beam was expanded using a concave lens with a focal length of −50 mm to avoid nonlinearity of the photodiode (PD) (Hamamatsu S3584-09). The reference signal was subtracted from the NV fluorescence signal using a custom-built transimpedance amplifier (TIA). This TIA consisted of a 1-kΩ resistor, a 47-pF capacitor, and an op-amp (ADA4898-1). Variable resistors were inserted between the two photodiodes to fine-tune the balance by changing the current phase.[16] Although an auto-balanced TIA[17] is also effective for pulse measurements, manual balancing proved more effective for suppressing laser intensity noise in this setup. The output from the TIA was AC-coupled, filtered, and amplified by a factor of 5 using an analog filter (SR560). We employed lock-in detection to mitigate 1/f and low-frequency laser intensity noise. A lock-in amplifier demodulated the TIA signal at 20 kHz with an additional gain of 10 to exceed the noise floor of the data acquisition system. The total transimpedance gain was $5 \times 10^4$ V/A. Accordingly, the system was able to reach the photon shot noise limit even with large photocurrents exceeding 10 mA.

The NV center spins were driven by a PCB MW antenna[18] comprising two parallel coupled lines with an inverted T-shape[19] with two ports for driving both frequencies of $|m_s = \pm 1\rangle$ to avoid the 3 dB loss associated with using an MW splitter. The antenna was designed to have a resonance frequency of approximately 3 GHz and an $S_{11}$ of less than –10 dB over a 200-MHz bandwidth. MW signals were generated and IQ-modulated with two signal generators (Keysight N5172B).

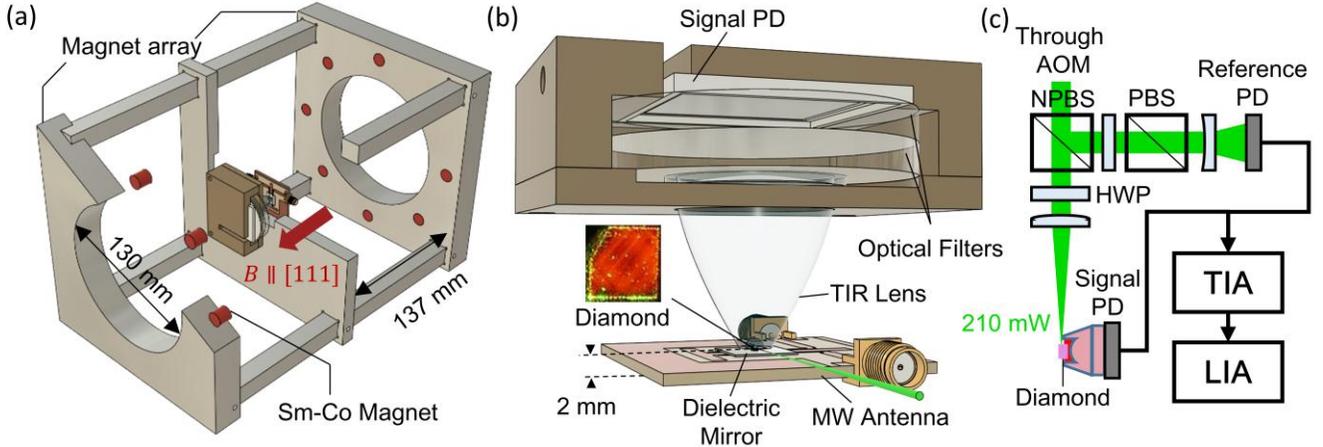

Fig. 1 Sensor setup. (a) Diamond quantum sensor in the magnet array system. A homogeneous bias magnetic field of 0.52 mT was applied along the [111] NV axis. This system was placed inside a three-layer magnetic shield. (b) Close-up view of the sensor head. The minimum sensor-to-sample distance was approximately 2.0 mm, which was limited by the thickness of the MW antenna and the dielectric mirror. A 210-mW, 532-nm laser beam was incident on the diamond. The inset shows multiple internal laser-beam reflections inside the diamond. (c) Schematic of optical setup and detection electronics.

Figure 2 (a) shows a pulsed-ODMR spectrum with four $^{14}$N-hyperfine triplets. The two outer triplets correspond to the [111] NV axis aligned along the bias magnetic field, whereas the two inner triplets correspond to the other three NV axes. The polarization of the laser beam was finely tuned to increase the amplitude of outer peaks. To suppress temperature variations induced by laser irradiation, the duration of the optical pulse for the initialization and read-out was kept constant at $20~\mu s$ for all sequences. From fitting, we approximately determined the resonance frequencies of both states $|m_s = \pm 1, m_I = 0\rangle$.

We applied the Ramsey protocol to estimate the dephasing time $T_2^*$ for both the SQ and DQ bases (Fig. 2(b)). To observe the fringes clearly, a differential detuning $\delta$ of 5 MHz was set from the $m_I = 0$ resonance frequency. From fitting, $T_2^*$ was estimated as 5.5 $\mu$s for the SQ basis and 4.4 $\mu$s for the DQ basis. The measured $T_2^*$ for the SQ basis was shorter than the predicted value of 8(1) $\mu$s limited by P1 centers.[20] We attribute this discrepancy to additional dephasing sources arising from inhomogeneity of crystal strain and temperature fluctuations in the diamond. Although the common-mode dephasing sources can be largely suppressed in the DQ basis, the theoretical P1-limited $T_2^*$ is reduced to 4.0(5) $\mu$s due to the doubled effective gyromagnetic ratio.[21] The experimentally obtained $T_2^*$ of 4.4 $\mu$s agreed with this value to within the experimental uncertainty, indicating that the DQ coherence is limited by P1 centers.

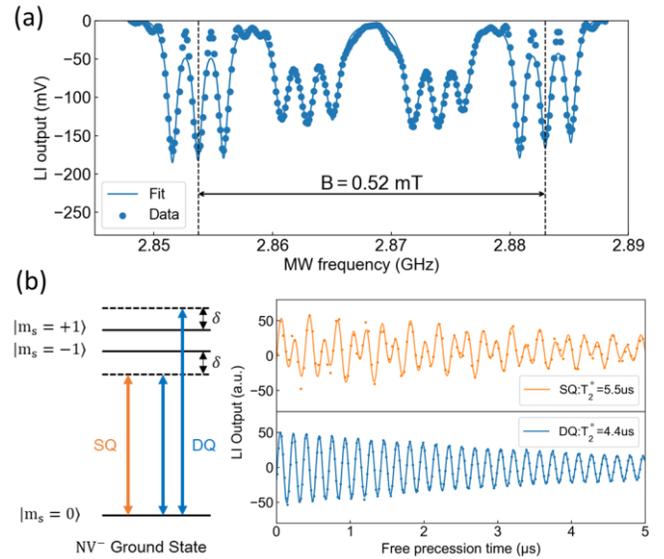

Fig. 2. (a) A pulsed-ODMR spectrum with a bias magnetic field of approximately 0.52 mT. (b) Ramsey fringes measured with a detuning of $\delta$ = 5 MHz from the $m_I = 0$ resonance for both SQ and DQ bases.

We employed the DQ 4-Ramsey sequence[21] with lock-in detection as a sensing scheme to reduce common-mode noise as shown in Fig. 3 (a). The lock-in reference pulse was synchronized with the laser and MW pulses, and its period was set to half the total sequence duration. The free precession time $\tau = 3.957~\mu s$ was chosen to maximize the slope of the Ramsey signal response. The duration of the laser irradiation was set to 20 μs, which was limited by the remaining time after subtracting $\tau$ and the MW pulse durations from half the lock-in modulation period. Figure 3 (b) shows the DQ Ramsey signal response as a function of the differential-mode detuning $\delta$. The experimentally obtained maximum slope of 761 pA/Hz was reasonably consistent with the calculated value of 802 pA/Hz using equation:[22]

$$\frac{dI}{d\delta} \approx 2\pi\tau\Delta m_s I C e^{-\left(\frac{\tau}{T_2^*}\right)^p}, \quad (1)$$

where $\Delta m_s = 2$, $\tau = 3.957~\mu s$, photocurrent $I = 5$ mA, Rabi contrast $C = 0.89$ %, $p=1.0$, and $T_2^* = 3.9~\mu s$ in the measurements in Fig. 3. The term for the number of readouts is omitted because lock-in detection was employed. The MW frequencies were fixed at the point of maximum slope (the zero-crossing point) to record the magnetic field signal.

Figure 3 (c) shows the positive frequency region of the double-sided amplitude spectral density (ASD) of the measured noise. The orange spectrum corresponds to the magnetically sensitive configuration with on-resonance MW frequencies, while the blue spectrum corresponds to the magnetically insensitive configuration at off-resonance MW frequencies (2.5 and 3.0 GHz). The solid black line represents the photon-shot-noise-limited sensitivity of 1.9 pT/√Hz, calculated using eq. (2):[9]

$$\eta_{Ramsey}^{ensemble,shot} = \frac{\sqrt{2qI}}{\gamma_e \left|\frac{dI}{d\delta}\right|_{exp}}, \quad (2)$$

where $q = 1.6 \times 10^{-19}$ C, $\gamma_e = 28 \times 10^9$ Hz/T and $\left|\frac{dI}{d\delta}\right|_{exp} = 761$ pA/Hz. The $\sqrt{2}$ factor accounts for uncorrelated shot noise due to the balanced detection. The noise in the magnetically insensitive configuration was very close to the photon shot noise limit. In the magnetically sensitive configuration, the noise slightly increased to 2.93(7) pT/√Hz in the frequency range of 100–400 Hz. The 50-Hz harmonics originate from power line interference coupled through the MW-related components. The increased noise at frequencies below 100 Hz is likely attributed to MW phase noise. The MW phase noise will be suppressed by employing a dual-tone frequency mixing configuration[23] or using an ultralow phase-noise signal generator.

Notably, a previous work[7] employed all four NV orientations, which enhanced the effective sensitivity by a factor of $4/\sqrt{3} \approx 2.3$ due to the exact degeneracy of all four crystallographic orientations. Taking this difference into account, the sensitivity achieved in this study can be regarded as comparable at low frequencies of 100–200 Hz. Furthermore, biomagnetic fields typically decay as $r^{-2}$ or faster with the distance.[3] Consequently, reducing the sensor-to-sample distance by a factor of 2 is equivalent to improving the sensitivity by a factor of 4 or higher. This characteristic highlights the suitability of our quantum sensor based on the Ramsey method to measure biomagnetic fields. In addition, the sensitivity demonstrated here exceeds that of previously reported CW-ODMR-based biomagnetic field sensors that do not employ magnetic flux concentrators.[2,3,5]

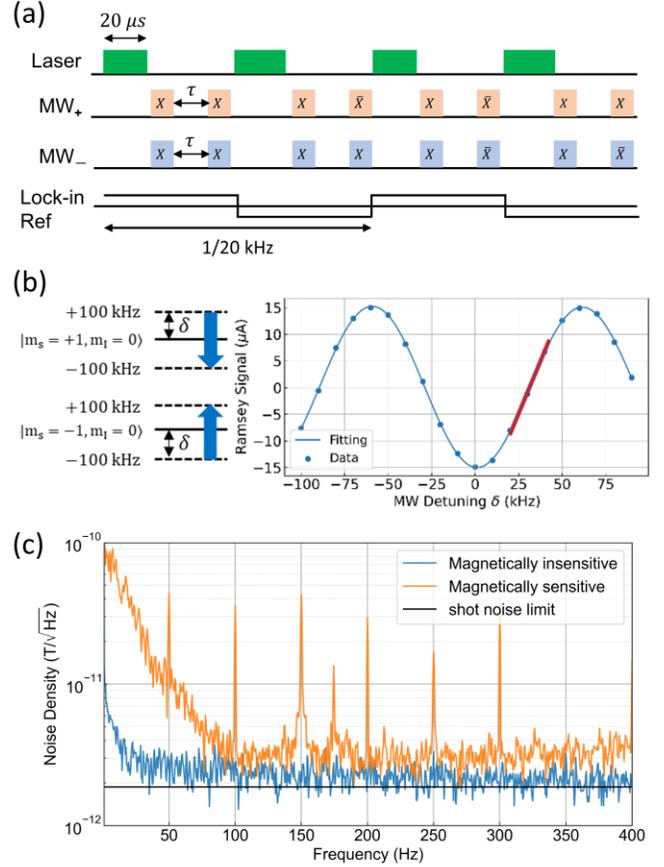

Fig. 3. (a) DQ 4-Ramsey sequence using lock-in detection. The pulse widths and intervals are not drawn to scale. (b) DQ 4-Ramsey signal response as a function of MW detuning. The slope at the zero-crossing was 761 pA/Hz, as indicated by a red line. (c) Sensitivity evaluation. The orange spectrum corresponds to the magnetically sensitive configuration with on-resonance MW fields. The blue spectrum corresponds to the magnetically insensitive configuration using off-resonance MW fields. The displayed frequency range was limited to below 400 Hz due to the bandwidth of the lock-in amplifier.

Finally, to demonstrate the sensor's performance at a short sensor-to-sample distance, we measured a magnetic signal using a dry phantom designed to mimic an MEG signal.[24] This phantom comprises an isosceles-triangle current path. The base of the triangle, which mimics an equivalent current dipole (ECD), was positioned at a distance of approximately 2.5 mm from the diamond (Fig. 4 (a)). An AC current of 50 $\mu$A at a frequency of 77 Hz was applied to the dry phantom. The phantom generated a magnetic field that followed the Sarvas formula[25]

$$\boldsymbol{B}(\boldsymbol{r}) = \frac{\mu_0}{4\pi F^2}[F\boldsymbol{Q} \times \boldsymbol{r_0} - \{(\boldsymbol{Q} \times \boldsymbol{r_0}) \cdot \boldsymbol{r}\}\nabla F], \quad (3)$$

where

$$F = |\boldsymbol{r} - \boldsymbol{r_0}|(|\boldsymbol{r} - \boldsymbol{r_0}||\boldsymbol{r}| + |\boldsymbol{r}|^2 - \boldsymbol{r_0} \cdot \boldsymbol{r}). \quad (4)$$

The graph on the right side in Fig. 4 (a) shows a simulated distribution of the magnetic field $B_z$ calculated using Eq. (3), with numerical parameters for a single ECD $\boldsymbol{Q} = (0, 35, 0)$ nA·m, the center position of the sensor $\boldsymbol{r} = (x, y, 12\text{ mm})$ and the position of the ECD $\boldsymbol{r_0} = (0, 0, 9.5)$ mm. This simulation represents the effective magnetic field sensed by the diamond, which was obtained by spatially averaging the field over the 0.9 mm × 0.9 mm sensing area. Depending on the sensor's center position $(x, y)$ within the assumed scan area (2.0 mm × 2.0 mm), the magnetic field amplitude was predicted to vary within ±150 pT. As shown in Fig.4 (b), the experimental signal from the phantom was detected without averaging after applying a digital filter. The peak amplitude determined by fitting the data was 77.7(2) pT. In the same setup without AC current, the double-sided ASD was 14.3 pT/$\sqrt{\text{Hz}}$ at 77 Hz. Considering the contribution from both sidebands and a 0.8-Hz-bandwidth digital filter, the RMS noise amplitude was calculated to be 18.1 pT, yielding a signal-to-noise ratio (SNR) of $77.7/18.1 \approx 4.3$. For comparison, a state-of-the-art Ramsey magnetometer[6] exhibited a superior noise floor of about 2 pT/$\sqrt{\text{Hz}}$ at 77 Hz. However, due to its sensor configuration, the sensor-to-sample distance would be expanded to 7.1 mm (6.6 mm sensor gap[7] + 0.5 mm phantom depth). According to the Sarvas formula[25], the magnetic signal from the same phantom source would be reduced to ≈9.6 pT by a geometrical factor of $(7.1/2.5)^2$. Assuming the same filter condition as ours, the RMS noise for this sensor would be 2.5 pT. Consequently, the SNR is estimated to be ≈ 3.8, similar to our SNR of 4.3. Thus, the magnetometer developed in this study is confirmed to have a high detectability of a spatially confined weak current-dipole signal similar to that of the state-of-the-art magnetometer,[7] as the magnetic-field sensitivity difference is compensated by the sensor-to-sample distance.

Further sensitivity improvements are expected by eliminating the contribution of photon shot noise from the reference beam, which would provide an additional factor of $\sqrt{2}$ enhancement.[7,17] In addition, a longer $T_2^* = 15.8\ \mu$s has been reported[9] for low-nitrogen diamond without employing spin-bath driving,[26] corresponding to a further improvement by a factor of 3.6. By combining these approaches, a photon shot noise limit of 0.37 pT/$\sqrt{\text{Hz}}$ might appear achievable. However, decreasing the concentration of nitrogen for longer $T_2^*$ would make charge-state conversion from the negatively charged NV$^-$ to neutral NV$^0$ more likely, and hence lead to reduction of the number of sensing NV$^-$ centers. Therefore, we expect that preserving the product of the photocurrent and contrast, $I \times C$, through active charge-state stabilization techniques, such as optical initialization protocols[27,28] or the application of electric fields[29], while employing low-nitrogen diamond samples will lead to sub-picotesla biomagnetic field diamond magnetometers with a short sensor-to-sample distance.

In conclusion, we developed a highly sensitive Ramsey diamond magnetometer with a short sensor-to-sample distance for biomagnetic measurements. By employing the light-trapping diamond waveguide technique and a TIR lens, we achieved efficient fluorescence collection with a low laser power of 210 mW and thus mitigated the heating issues associated with conventional high-sensitivity setups. The sensor demonstrated a magnetic field sensitivity of 2.93(7) pT/$\sqrt{\text{Hz}}$ in the 100-400 Hz range using a DQ 4-Ramsey sequence with lock-in detection. Furthermore, we successfully detected a weak magnetic signal of 77.7(2) pT from a dry phantom at a sensor-to-sample distance of 2.5 mm without signal averaging, with a sensing diamond volume of 0.28 mm³. This short distance measurement, enabled by our low-thermal-load and compact design, prevents severe spatial signal attenuation and yields a high effective sensitivity for real-time signal detection. These results represent a significant step toward practical MEG and other weak biomagnetic field applications. Future improvements in diamond crystal quality to extend $T_2^*$ and the implementation of advanced noise suppression techniques are expected to push the sensitivity of the sensor system into the sub-picotesla regime, which will open new avenues for noninvasive biomagnetic field sensing.

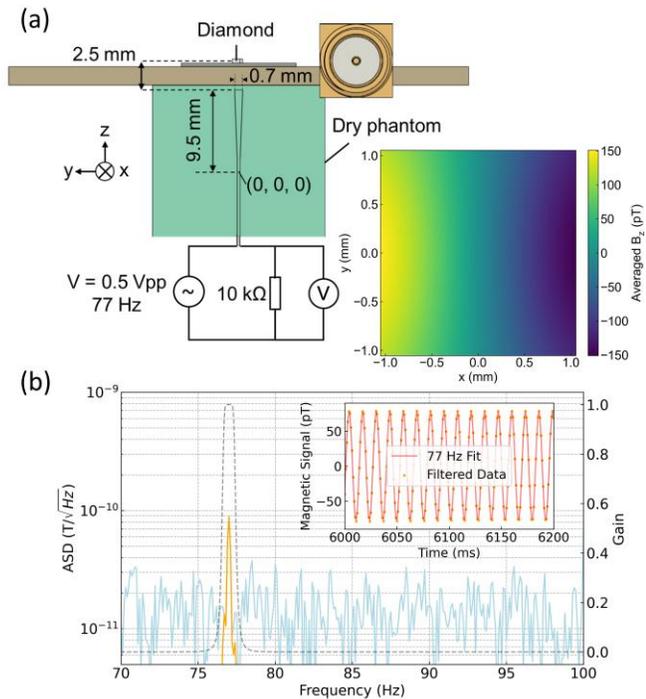

Fig. 4. (a) Schematic of the experimental setup with a dry phantom, omitting bias magnets and the photon collecting system for clarity. The right graph shows a simulated magnetic-field map generated by the dry phantom, obtained by spatially averaging the magnetic field over the diamond sensing area (0.9 mm × 0.9 mm) and scanning this averaging window across the phantom. (b) The main panel displays the measured noise spectrum (light blue) and digitally filtered magnetic field signal (orange) and the digital filter gain (gray dashed line). The inset panel shows the digital filtered signal and fitted line which shows the amplitude of 77.7(2) pT at a frequency of 77 Hz.

ACKNOWLEDGMENTS

This work was supported by Q-LEAP Grant Numbers JPMXS0118067395 and JPMXS0118068379, JST ASPIRE Grant Number JPMJAP24C1, and CSTI SIP "Promoting the application of advanced quantum technology platforms to social issues".